\documentclass[runningheads]{llncs}
\usepackage{graphicx}
\usepackage{amsmath}
\usepackage{booktabs}
\usepackage{multirow}
\usepackage{hyperref}
\usepackage{float}
\usepackage{array}
\usepackage{xcolor}
\usepackage{cite} 
\usepackage{caption}

\begin{document}

\title{Comparative Analysis of Large Language Models in Generating Telugu Responses for Maternal Health Queries}
\titlerunning{LLMs Evaluation - Maternal Health Queries in Regional Languages}
\author{Anagani Bhanusree\inst{1} \and
Sai Divya Vissamsetty\inst{1} \and
K VenkataKrishna Rao\inst{1} \and
Rimjhim\inst{1}}
\authorrunning{A Bhanusree et al.}

\institute{National Institute of Technology, Warangal, India\\
\email{ab25csr1p11@student.nitw.ac.in,vs25csr1p10@student.nitw.ac.in, kvkrao@nitw.ac.in, rimjhim@nitw.ac.in}}
\maketitle
\vspace{-0.25em} 

\begin{abstract}
Large Language Models (LLMs) have been progressively exhibiting there capabilities in various areas of research. The performance of the LLMs in acute maternal healthcare area, predominantly in low resource languages like Telugu, Hindi, Tamil, Urdu etc are still unstudied. This study presents how ChatGPT-4o, GeminiAI, and Perplexity AI respond to pregnancy related questions asked in different languages. A bilingual dataset is used to obtain results by applying the semantic similarity metrics (BERT Score) and expert assessments from expertise gynecologists. Multiple parameters like accuracy, fluency, relevance, coherence and completeness are taken into consideration by the gynecologists to rate the responses generated by the LLMs. Gemini excels in other LLMs in terms of producing accurate and coherent pregnancy relevant responses in Telugu, while Perplexity demonstrated well when the prompts were in Telugu.  ChatGPT's performance can be improved. The results states that both selecting an LLM and prompting language plays a crucial role in retrieving the information. Altogether, we emphasize for the improvement of LLMs assistance in regional languages for healthcare purposes.

\keywords{Large Language Models \and Maternal Health \and BERT Score \and Indic Languages \and Low Resource Languages.}
\end{abstract}

\section{Introduction}
\vspace{-3mm}
Advancements in LLMs have greatly expanded their applicability across a variety of languages and specialized fields. However, their capabilities differ significantly when it comes to low resource languages such as Telugu, Hindi, Tamil, Urdu etc. While some models excel in generating fluent and contextually suitable text, others demonstrate stronger abilities in providing precise, domain specific information. The rapid evolution of LLM architectures and the diversity of their training data have led to inconsistent performance across Indic languages. This inconsistency highlights the necessity for comprehensive evaluation methods that account for linguistic nuances. Furthermore, deploying LLMs in sensitive areas like maternal healthcare demands an emphasis on accuracy and trustworthiness. This research seeks to fill these gaps by systematically examining how well, different LLMs understand and generate Telugu language content related to maternal health, considering both bilingual and domain specific factors.

Previous studies have explored the accuracy and linguistic capabilities of various LLMs for pregnancy related queries in Telugu and other regional languages. For example, Kishore et al. \cite{kishore2024evaluating} found Gemini to have superior grammatical and idiomatic skills, while ChatGPT showed higher factual correctness. Vaidya et al. \cite{vaidya2025analysis} highlighted the strong performance of several LLMs in widely spoken Indic languages, although challenges remain for low resource languages such as Telugu. Lima et al. \cite{lima2025quality} evaluated LLMs in providing maternal health information in resource-constrained environments, noting both their potential and limitations. 
Lee et al. \cite{lee2025readability} noted readability gaps between ChatGPT
and Bard in labor epidural FAQs. Collectively, these works emphasize the need
for rigorous evaluation, contextual adaptation, and cautious clinical integration
of LLMs, especially in underserved regions where digital tools can greatly enhance healthcare access. 

Other researchers, including Chalamalasetti et al. \cite{kranti2025mata} introduced MATA, a benchmark suite for Telugu that assesses various linguistic phenomena and highlights challenges in relying solely on automated metrics. Singh et al. \cite{singh2024indicgenbench} introduce IndicGenBench, a comprehensive multilingual benchmark designed to assess the text generation capabilities of large language models across diverse Indic languages, emphasizing linguistic diversity and cross-lingual generalization. The study provides critical insights into existing models’ limitations and promotes advancement in Indic NLP evaluation frameworks. Lastly, Aggarwal et al. \cite{aggarwal2022indicxnli} introduced IndicXNLI, a benchmark dataset for evaluating multilingual natural language inference across 11 Indian languages, enabling assessment of cross-lingual transfer in models like mBERT, XLM-R, IndicBERT, and MuRIL.

\vspace{-3mm}

\section{Data Collection}
\vspace{-1mm}
Here, we formulated a collection of regular pregnancy related questions reflecting typical concerns raised by expectant and lactating mothers. To capture linguistic diversity, these questions were prepared in both English and Telugu. The topics covered, ranged from nutrition, symptom management to fetal development and antenatal care. Each question was presented twice to the LLMs, first in English and then in Telugu. The LLMs were instructed to provide answers in Telugu, aiming to mimic the responses a practicing gynecologist fluent in Telugu might offer. Additionally, a notable expert gynecologists generated reference answers in Telugu to serve as a benchmark for evaluation. This bilingual question and answer dataset forms the foundation for assessing the LLMs performance.
\vspace{-0.25em} 

\section{Methodology}
Our evaluation approach combines automated semantic analysis with detailed expert review. Initially, we measured the semantic similarity between the LLM generated answers and the expert responses using BERT Score, a metric that utilizes contextual embeddings to capture meaning beyond exact word matches. This allowed an objective comparison of how closely the LLMs outputs aligned with medically accurate information.

Simultaneously, we involved ten practicing gynecologists fluent in Telugu to perform a qualitative evaluation. These experts assessed each answer based on five criteria: accuracy (correctness of medical facts), fluency (grammatical and natural use of Telugu), relevance (appropriateness and focus on the query), coherence (logical structure and flow), and completeness (coverage of the question’s aspects). Each criterion was rated on a Likert scale from 1 (lowest) to 10 (highest).

By integrating both quantitative semantic scores and qualitative expert judgments, our methodology captures the multifaceted nature of LLMs performance. We also investigated how the language of the input prompt English versus Telugu affects the quality of Telugu responses. This comprehensive evaluation framework provides valuable insights into the practical utility of LLMs for healthcare communication in regional languages.

\section{Results and Discussion}
\vspace{-3mm}

Throughout the results and analysis, we use the following abbreviations for clarity: D represents the expert doctor’s reference responses. The abbreviations EP and TP denote responses generated by Perplexity when prompted in English and Telugu, respectively. Similarly, EG and TG correspond to Gemini’s responses for English and Telugu prompts. Lastly, EC and TC indicate ChatGPT’s answers to English and Telugu questions, respectively. These labels are consistently used to facilitate comparison across models and prompt languages.
\vspace{3mm}

\textbf{A) Semantic Similarity Analysis Using BERT Score:}

To objectively assess how closely the responses generated by the LLMs match to the expert gynecologists’ answers, we employed BERT Score, which evaluates semantic similarity using contextual embeddings. Table ~\ref{tab:bert_score}, summarizes the average precision, recall, and F1 scores for each model’s outputs based on the language of the prompt.

Among all configurations, Perplexity prompted in English (EP) achieved the highest F1 score of 0.704, indicating a strong semantic alignment with expert responses. This suggests that while the model’s training and architecture enable it to grasp medical context effectively, the choice of prompt language can subtly influence the outcome.
While, Gemini’s responses, whether prompted in English (EG) or Telugu (TG), also demonstrated high semantic similarity, with scores closely trailing Perplexity’s best result.
ChatGPT showed a slightly lower semantic overlap, particularly when prompted in English (EC), receiving the lowest F1 score of 0.684. 
This reflects a moderate but consistent gap in capturing all nuances of maternal healthcare and linguistic expression in Telugu.

\vspace{3mm}
\begin{table}[htb]
\centering
\renewcommand{\arraystretch}{1.3} 
\setlength{\tabcolsep}{5pt} 
\caption{BERT-Score Comparison between Gynecologist Responses and LLM Responses}
\label{tab:bert_score}
\begin{tabular}{lccc}
\toprule
\textbf{Comparison} & \textbf{Precision (Mean)} & \textbf{Recall (Mean)} & \textbf{F1 Score (Mean)} \\
\midrule
D vs EP & 0.692 & 0.718 & 0.704 \\
D vs TP & 0.688 & 0.719 & 0.703 \\
D vs EG & 0.694 & 0.691 & 0.693 \\
D vs TG & 0.689 & 0.691 & 0.690 \\
D vs EC & 0.662 & 0.708 & 0.684 \\
D vs TC & 0.685 & 0.720 & 0.702 \\
\bottomrule
\end{tabular}
\end{table}
 Interestingly, when prompted in Telugu (TC), ChatGPT’s semantic alignment improved, highlighting the model’s sensitivity to input language in generating regional language outputs.
These findings underscore that while semantic similarity provides a useful quantitative benchmark, it does not fully capture the maternal knowledge appropriateness or linguistic quality of generated responses, necessitating deeper expert evaluation.

\vspace{5mm}
\textbf{B) Expert Evaluation of LLMs Generated Responses:}

\begin{figure}
    \centering
    \includegraphics[width=1\linewidth]{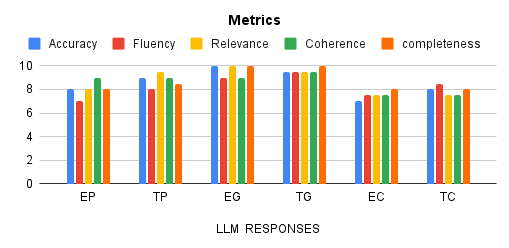}
    \caption{Bar graph representing the Average Scores of Evaluation Metrics}
    \label{fig:placeholder}
\end{figure}

Ten practicing gynecologists fluent in Telugu assessed the responses on five critical parameters: accuracy, fluency, relevance, coherence, and completeness. Each criterion was rated on a scale from 1 (poor) to 10 (excellent), and the average scores are presented in Table ~\ref{table:expert_evaluation}.
Gemini consistently received the highest ratings across all five metrics, especially for responses generated from English prompts. Its answers were noted for their medical precision, logical flow, and thoroughness, often matching or exceeding the quality of expert reference responses. Fig: ~\ref{fig:placeholder} indicates Gemini’s robust understanding of domain specific knowledge combined with strong language proficiency.

\vspace{-8mm}
\begin{table}[H]
\centering
\caption{Average Expert evaluation scores for LLM generated responses}
\begin{tabular}{c c c c c c}
\hline
\textbf{LLM Responses} & \textbf{Accuracy} & \textbf{Fluency} & \textbf{Relevance} & \textbf{Coherence} & \textbf{Completeness} \\
\hline
EP & 8.0 & 7.0 & 8.0 & 9.0 & 8.0 \\
TP & 9.0 & 8.0 & 9.5 & 9.0 & 8.5 \\
EG & 10.0 & 9.0 & 10.0 & 9.0 & 10.0 \\
TG & 9.5 & 9.5 & 9.5 & 9.5 & 10.0 \\
EC & 7.0 & 7.5 & 7.5 & 7.5 & 8.0 \\
TC & 8.0 & 8.5 & 7.5 & 7.5 & 8.0 \\
\hline
\end{tabular}
\label{table:expert_evaluation}
\end{table}

\vspace{-2 em}
Perplexity showed a marked improvement when prompted in Telugu, achieving higher fluency and relevance scores compared to English prompts. This suggests that the model leverages Telugu input effectively to produce more contextually appropriate and linguistically natural replies.
ChatGPT responses were generally rated as moderate. Although the fluency was acceptable, the model occasionally missed some critical maternal healthcare details, resulting in lower accuracy and completeness scores. The performance gap between English and Telugu prompts for ChatGPT was less pronounced but still noticeable, with Telugu prompts yielding marginally better results.

\textbf{C) Additional Observations and Key Insights:}
 The study clearly shows that the prompt language plays a prominent role in the quality of Telugu responses, with Perplexity improving when given Telugu prompts. Gemini maintains strong results regardless of prompt language, likely due to its architecture and training data. Automated metrics like BERT Score offer a useful baseline but miss subtle maternal health and linguistic nuances captured by expert reviews. 
 
 ChatGPT’s moderate performance points to the need for further fine tuning for regional medical contexts. The gap between semantic similarity and expert judgment highlights the importance of combining both evaluation methods. Selecting the right LLM and prompt style is critical for delivering trustworthy healthcare information in regional languages. These findings emphasize how model choice and prompt design affect AI reliability in maternal health communication.

\section{Conclusion and Future Work}
Here, we examined how different LLMs respond to maternal health questions in Telugu, revealing that both the choice of model and the language used to prompt it have a big impact on the quality of answers. Gemini stood out by giving clear and accurate replies consistently, while Perplexity’s responses got better when prompted in Telugu. ChatGPT showed average results but could improve with more tuning. Moving forward, we plan to include a wider variety of maternal healthcare questions and cover more regional languages to make the models more robust. We’re also interested in fine tuning these models specifically for maternal healthcare and understanding how people trust and use AI generated advice. These efforts will help make AI tools more dependable and acceptable in real healthcare settings. Ultimately, picking the right model and crafting prompts carefully are key to trustworthy AI support in maternal health.
\bibliographystyle{IEEEtran}    
\bibliography{reference} 
\end{document}